\documentclass[iop,apj]{emulateapj}

\usepackage{apjfonts}
\usepackage{amssymb,amsmath}
\usepackage{hyperref}

\def\hst{{\it HST}}

\newcommand{\kms}{\>{\rm km}\,{\rm s}^{-1}}
\newcommand{\masyr}{\>{\rm mas}\,{\rm yr}^{-1}}
\newcommand{\kpc}{\>{\rm kpc}}

\newcommand{\muw}{\mu_{W}}
\newcommand{\mun}{\mu_{N}}

\shorttitle{{\it HST} Proper Motions of Stars in Stellar Streams}
\shortauthors{Sohn et al.}

\begin{document}

\title{{\it Hubble Space Telescope} Proper Motions of Individual Stars in Stellar Streams: \\
Orphan, Sagittarius, Lethe, and the New ``Parallel'' Stream}

\author{Sangmo Tony Sohn\altaffilmark{1,2}, 
Roeland P. van der Marel\altaffilmark{2}, 
Nitya Kallivayalil\altaffilmark{3},
Steven R. Majewski\altaffilmark{3},
Gurtina Besla\altaffilmark{4},
Jeffrey L. Carlin\altaffilmark{5},
David R. Law\altaffilmark{2},
Michael H. Siegel\altaffilmark{6},
and
Jay Anderson\altaffilmark{2}
}

\altaffiltext{1}{Department of Physics and Astronomy, The Johns Hopkins University, 
                 Baltimore, MD 21218, USA}
\altaffiltext{2}{Space Telescope Science Institute, 
                 3700 San Martin Drive, Baltimore, MD 21218, USA}
\altaffiltext{3}{Department of Astronomy, 
                 University of Virginia, 
                 Charlottesville, VA 22904-4325, USA}
\altaffiltext{4}{Steward Observatory, University of Arizona,
                 933 North Cherry Avenue, Tucson, AZ 85721, USA}
\altaffiltext{5}{LSST, 
                 933 North Cherry Avenue, Tucson, AZ 85721, USA}
\altaffiltext{6}{Department of Astronomy, 
                 Pennsylvania State University,
                 525 Davey Laboratory, University Park, PA 16802, USA}

\email{tsohn@jhu.edu}

\begin{abstract}
We present a multi-epoch {\it Hubble Space Telescope} (\hst) study of 
stellar proper motions (PMs) for four fields along the Orphan Stream.
We determine absolute PMs of several individual stars per target field 
using established techniques that utilize distant background galaxies 
to define a stationary reference frame. Five Orphan Stream stars are 
identified in one of the four fields based on combined color-magnitude 
and PM information. The average PM is consistent with the existing model 
of the Orphan stream by Newberg et al. In addition to the Orphan stream 
stars, we detect stars that likely belong to other stellar streams.
To identify which stellar streams these stars belong to, we examine 
the 2-d bulk motion of each group of stars on the sky by subtracting 
the PM contribution of the solar motion (which is a function of  
position on the sky and distance) from the observed PMs, and comparing 
the vector of net motion with the spatial extent of known stellar streams. 
By doing this, we identify candidate stars in the Sagittarius and Lethe 
streams, and a newly-found stellar stream at a distance of $\sim 17$ kpc, 
which we tentatively name the ``Parallel stream''. Together with our 
Sagittarius stream study (Sohn et al., 2015, ApJ, 803, 56), this work 
demonstrates that even in the {\it Gaia} era, \hst\ will continue to be 
advantageous in measuring PMs of old stellar populations on a star-by-star 
basis, especially for distances beyond $\sim 10$ kpc.
\end{abstract}

\keywords{astrometry ---
Galaxy: halo ---
Galaxy: kinematics and dynamics ---
proper motions}

\section{Introduction}
\label{sec:intro}

Stellar streams in the Milky Way (MW) halo are direct evidence of the 
hierarchical nature of cold dark matter-based structure formation. 
They prove that our Galaxy has grown, at least in part, via the tidal 
disruption of dwarf galaxies and globular clusters. Discovering and 
characterizing tidal streams is therefore a critical test for 
structure formation models. On global scales, streams are powerful 
probes of radial profile and shape of the MW's dark matter halo 
\citep[e.g.,][]{joh99,pen12}. Because they are kinematically cold, 
streams are also emerging as useful tools for constraining small-scale 
dark matter structures: direct impacts from subhalos may dynamically 
heat a stream and create gaps in surface density along the debris
\citep{joh02,yoo11,car12a,car12b,erk15a,erk15b,bov16}.

The structure of a tidal stream dictates the regimes over which it is 
sensitive as a dynamical probe. A long stream such as the Sagittarius 
(Sgr) Stream has been mapped a full 360$\degr$ on the sky and thus 
allows sensitivity to the global potential, but might have complicated 
dynamics because it is influenced by a comparatively large progenitor. 
By contrast, thin streams, such as those belonging to globular clusters 
like Pal 5, are sensitive to small-scale lumpiness, but might suffer 
from limitations in their phase-space coverage and hence from 
degeneracies in fitting different orbital parameters to a limited set 
of observables \citep[e.g.,][]{kop10}. In general, the parameter space 
involved in modeling these interactions is large, and prominently 
lacking are proper motions (PMs) as a function of position along 
a stream.

Currently, only a few PM datasets exist for streams, including patches 
of the Sgr Stream \citep{car12,kop14,soh15}, and for the GD-1 Stream 
\citep{kop10}, which is unfortunately very close in and probes the 
Galactic disk rather than the halo. The Sgr Stream is the brightest and 
most prominent stellar stream in the MW and probes the halo. As a result 
it has been the the topic of intense data-collection and modeling efforts. 
In the past, no one model had been capable of simultaneously reproducing 
both the angular position and the distances/radial velocities of tidal 
debris in the Sgr leading arm in a static axisymmetric MW dark halo. 
\citet{law09} showed that the data can be reconciled by adopting a 
triaxial halo for the MW. The best fit halo model is then near oblate but 
with its symmetry axis perpendicular to that of the MW disk 
\citep{law10,deg13}. Although this work represents one of the most 
sophisticated data-model comparisons to date, its conclusions are quite 
surprising and has been the topic of intense debate \citep{deb13}. 
Moreover, in recent works uncovering new Sgr debris and tracing these to 
their furthest extents, the debris has been found to deviate significantly 
from the \citet{law10} model \citep{bel14,kop15}. In any case, 
PM measurements have enormous potential to better restrict both the 
progenitor orbit and the MW's gravitational potential \citep[e.g.,][]{kop10,car12}.

The astrometric capabilities available with the {\it Hubble Space Telescope} 
(\hst ) are an order of magnitude better than what can be achieved from 
the ground, so \hst\ has the potential to become a workhorse for measuring 
PMs along stellar streams in the MW halo. We have in fact demonstrated this 
to be the case in our Sgr Stream study of \citet{soh15}. While the low 
surface brightness of tidal streams, combined with the relatively small 
size of the \hst\ field of view, limited the detected number of Sgr stream 
stars to about a few dozen stars per field, the accurate PMs allowed a 
confident separation of stream stars from foreground (disk) stars. This 
enabled an accurate measurement of the average PM of the stream. Furthermore, 
the accurate PMs for individual stars also allowed us to identify multiple 
kinematical components within the same \hst\ fields. These discriminatory 
capabilities of \hst\ are crucial when constraining properties of the MW 
dark halo because PM measurements of streams affected by contamination from 
other kinematical components can lead to erroneous interpretations. In this 
study, we once again rely on the \hst\ to measure PMs of tidal stream stars, 
in this case those along the Orphan Stream.

The Orphan stream \citep{gri06b,bel06}, $\sim 60$ deg long and only $\sim 2$ 
deg ($\sim 0.7$ kpc) wide, is approximately five times narrower than the 
Sgr Stream while slightly closer to us, although a factor of $\sim 2$ lower 
in surface brightness. Given its atypical properties, the Orphan stream 
provides us with the opportunity to study the fainter end of the family of 
objects that built up the stellar halo of the MW. \citet{bel07} found a 
distance gradient along the stream, and also published radial velocities 
from sparse samples of SDSS data in two fields, one at the close end of 
the stream (at $\sim20$ kpc) and one at the distant end of the stream 
(at $\sim32$ kpc). Later, \citet{new10} published radial velocities for 
blue horizonal-branch (BHB) stars along the stream from SDSS/SEGUE, and 
\citet{cas13} and \citet{cas14} published K-giant spectra from 
AAT/AAOMEGA as well as high-resolution Magellan/IMACS follow up on their 
candidates. These latter studies find a significant metallicity dispersion 
and [$\alpha$/Fe] trends, which support the theory that the progenitor is a 
dwarf spheroidal satellite. \citet{fel07} constructed tidal disruption models 
assuming that the UMa II dSph is the progenitor of the stream, and concluding 
that some young halo globular clusters and the Complex~A gas cloud may also 
be associated. However, this study had trouble fitting the near-field velocity 
data. \citet{jin07} explored in more detail the possibility that Complex~A 
and the Orphan stream were related to each other, but they were unable to 
obtain a good fit under this assumption. \citet{sal08} and \citet{new10} 
both fitted orbits to the available data, and found that the orbit of 
the Orphan stream does not intersect either UMa~II or Complex~A. 
The data are consistent with a single wrap of the trailing arm of a 
fully disrupted satellite, or with a progenitor outside of the SDSS 
coverage area. More recently, \citet{gri15} detected the Orphan stream 
in the southern sky using Dark Energy Camera (DECam) observations 
extending the total known length of the Orphan stream to 108\degr.
They have also found an overdensity in the stream that seems consistent
with the Orphan progenitor. This finding combined with the dynamical model 
of \citet{new10} implies that the original detection of the stream is the 
leading arm.

In this paper, we present results from our \hst\ project to measure 
PMs of individual stars along the Orphan stream. The paper is organized 
as follows. In Section~2, we describe the data used for this study, 
and the determination of photometric and PM measurements for individual 
stars in our target fields. In Section~3, we describe the identification 
of Orphan stream stars in each of the fields based on the combined 
color-magnitude diagram (CMD) and PM information. We also discuss the 
detection of stars that possibly belong to other known stellar 
streams (e.g., the Sgr Stream). In Section~4, we explore the 2-d 
motions of stars on the sky based on our PM measurements, and discuss 
the nature of the known and newly discovered kinematical components. 
Finally, we summarize our results in Section~5. 

\section{Observations and Data Analysis}
\label{sec:obs_and_data}

%
\begin{figure}
\epsscale{1.15}
\plotone{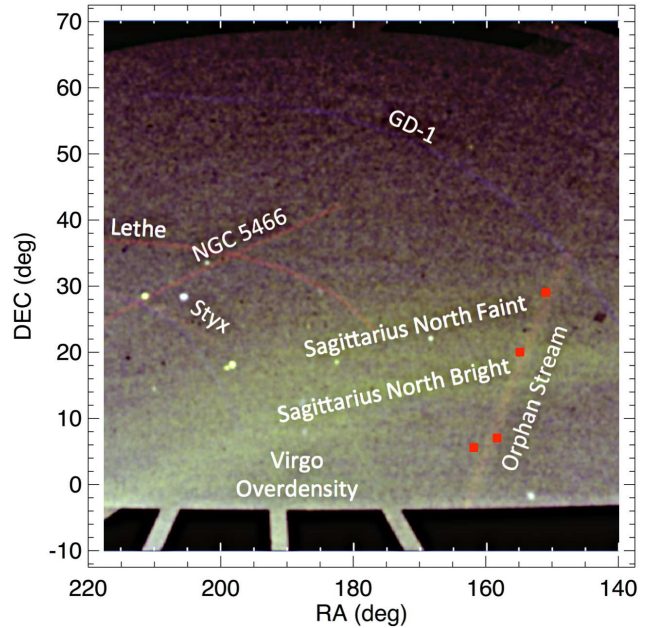}
\caption{Partial view of the SDSS footprint showing stellar streams
         at various distances in equatorial coordinates \citep[original 
         figure from][]{gri16}.
         The two bright branches that run roughly horizontal in this
         figure are the Sgr Stream: the bottom one is the main
         stream, and the top one is known as the ``faint arm.'' 
         The thin branch that runs roughly perpendicular 
         to the Sgr Stream is the Orphan Stream.
         Our four \hst\ target fields are overplotted as red squares.
         From top to bottom, these fields are denoted {\it ORPH-F1}, 
         {\it ORPH-F2}, {\it ORPH-F3}, and {\it ORPH-F4}.
         \label{fig:fields}
        }
\end{figure}
%

\subsection{First- and Second-epoch Data}
\label{sec:data}

The original goal of this project was to measure \hst\ PMs of stars 
located along the Orphan stream using \hst\ data. To achieve this 
goal, we searched the \hst\ archive for existing deep imaging 
serendipitously located in the vicinity of the Orphan stream.
We identified and selected four fields (Figure~\ref{fig:fields}) with 
characteristics as summarized in Table~\ref{tab:fields}. We note that 
fields {\it ORPH-F1} and {\it -F2} are located close to the densest 
parts of the stream \citep[$B_{\rm Orphan} \lesssim 0.3$ deg, where 
$B_{\rm Orphan}$ refers to the orthogonal distance to the Orphan Stream 
as defined by][]{new10}, whereas the other two fields 
({\it ORPH-F3} and {\it -F4}) are offset from the center of the 
stream by more than 1 deg in their closest points on the sky. 
Interestingly, two of our target fields happen to be also located on 
the two separated arms of the Sgr Stream: {\it ORPH-F1} is 
on the faint arm \citep[or branch~B as denoted by][]{bel06} and 
{\it ORPH-F2} is on the bright arm (or branch~A).

The first-epoch data of our target fields were observed for 
various \hst\ programs (see Table~\ref{tab:fields}). The second-epoch 
data were obtained through our \hst\ Program GO-13443 (PI: R. van der 
Marel). We targeted the four fields using the same observational 
setups (i.e., pointing, telescope orientation, detector, and filters) 
as the first-epoch observations. For astrometry, we observed with 
the ACS/WFC detector using F814W for {\it ORPH-F1} and {\it -F4}, 
and F775W for {\it ORPH-F2} and {\it -F3} to match the first-epoch data. 
Hereafter, we denote both of these filters as $I$-band unless there
is a need to make a distinction between them. During our second-epoch 
observations, we also obtained F606W exposures to obtain color 
information of stars in our target fields for constructing CMDs.

\subsection{PM Measurements and Photometry}
\label{sec:PMmeasurement}

PMs of stars in our target fields were measured using the same 
technique as we used in \citet{soh15} for measuring PMs of stars
along the Sgr Stream. In summary, we stacked high-resolution images 
using our deeper second-epoch observations, identified stars and 
background galaxies from the stacks, created background galaxy 
templates, measured positions of stars and galaxies on images in 
each epoch with the templates, and measured shifts in positions 
of stars with respect to the background galaxies between the two 
epochs. In addition to the PMs, we have also measured photometry 
for each star in our target fields. For this, we combined images 
for each field per filter using {\it astrodrizzle}, measured 
stars within aperture radii of 0.1 mas (4 ACS/WFC pixels), and 
carried out aperture corrections to infinity following the method 
provided by \citet{sir05}. We then calibrated the photometry 
to the ACS/WFC VEGAMAG system using the time-dependent zero points 
provided by the STScI webpage.
\footnote{\url{http://www.stsci.edu/hst/acs/analysis/zeropoints}.}

%
\begin{deluxetable*}{lcccccccccc}
\tablecolumns{11}
\tablewidth{0pc}
\tablecaption{\hst\ Target Fields and Observations
              \label{tab:fields}}
\tablehead{
\colhead{} & \colhead{R.A.} & \colhead{Decl.}  & \colhead{$\Lambda_{\rm Orphan}$\tablenotemark{a}} & \colhead{$B_{\rm Orphan}$\tablenotemark{a}} & \multicolumn{3}{c}{Epoch 1} & \colhead{} & \multicolumn{2}{c}{Epoch 2 (Prog. ID 13443)} \\
\cline{6-8} \cline{10-11} \\
\colhead{Target Fields} & \colhead{(J2000)} & \colhead{(J2000)} & \colhead{(deg)} & \colhead{(deg)} & \colhead{Prog. ID} & \colhead{Epoch} & \colhead{Exp. Time (s)\tablenotemark{b}} & \colhead{} & \colhead{Epoch} & \colhead{Exp. Time (s)\tablenotemark{b}} 
}
\startdata
{\it ORPH-F1} & 10:03:48.9 & $+$29:06:12.8 &    $-$11.96 &  $-$0.31 & 9468 & 2003.29 & 6000 & & 2014.23 & 7339 \\
{\it ORPH-F2} & 10:19:16.4 & $+$20:02:11.6 & \phn$-$2.23 &  $-$0.27 & 9575 & 2002.41 & 3100 & & 2014.24 & 7284 \\
{\it ORPH-F3} & 10:33:12.9 & $+$07:03:15.4 &   \phs11.11 & \phs1.20 & 9984 & 2003.90 & 2444 & & 2014.97 & 7072 \\
{\it ORPH-F4} & 10:47:26.8 & $+$05:35:22.4 &   \phs13.66 &  $-$1.61 & 9877 & 2004.36 & 3630 & & 2014.28 & 7234
\enddata
\tablenotetext{a}{Coordinate system of the Orphan stream as defined by \citet{new10}.}
\tablenotetext{b}{Total exposure time of the $I$-band observations used for astrometric analysis.}
\end{deluxetable*}
%

\section{Identification of Stars in the Orphan Stream}
\label{sec:identification}

Our target fields are expected to contain not only stars that belong to 
the Orphan stream, but also other stars in the foreground or background. 
We follow a similar method used in \citet{soh15} for identifying stars 
associated with the Orphan stream using the information provided by 
the CMDs and PM diagrams. Specifically, we first overlay fiducial 
isochrones developed by the Dartmouth Stellar Evolution Database 
\citep{dot08} on the CMDs that represent the Orphan stream stellar 
population, and select stars consistent with these isochrones.
The stars in the Orphan stream are known to be mainly old and metal-poor
with metallicities ranging from [Fe/H] $= -1.5$ to $-2.5$ according
to spectroscopic observations \citep{new10,ses13,cas13}. Therefore, 
we decided to overlay isochrones with an age of 12 Gyr and with three    
different metallicities ([Fe/H] $= -1.5$, $-2.0$, and $-2.5$) 
that encompass the spectroscopic metallicity range of most stars 
identified as Orphan stream stars in previous spectroscopic surveys.
\footnote{The choice of age for our isochrones is not important since,  
as will be shown, all of the stars we identify lie in the CMD fainter 
than the main sequence turnoff.}
For the distance to the Orphan stream at each field location, we 
adopted the heliocentric distance based on the orbital model by 
\citet{new10}. Their models were tailored to fit the observed that  
of BHB stars along the Orphan stream and, because \citet{ses13} finds 
that \citet{new10}'s Model~5 fits the distance of RRab stars fairly well, 
that was our model of choice. Finally, we applied reddening to the 
isochrones based on the $E(B-V)$ values estimated from interpolating 
the reddening maps of \citet{sch98}. The total absorption values were 
then adopted from Table~6 of \citet{sch11}. We note that the foreground 
reddening of our target field is generally very low, with $E(B-V)$ values 
in the range 0.019--0.022. The isochrones are plotted in black lines in 
the CMD of each field. Stars are considered to be consistent with an 
Orphan-like population if they lie on one of the isochrones 
(or in between two of the isochrones) within their color errors.
We note that the exact choice for the stream's stellar population
properties, reddening, or distance are not critical for this study 
as our main purpose is to determine which stars in each observed 
CMD are reasonably associated with the Orphan stream. 

Once candidate member stars are selected based on their location in the 
CMDs, we inspect the PM diagrams to locate tight clumps that represent 
kinematically-cold groups of stars. If such clumps exist, they are 
compared to the PM predictions provided by \citet{new10} to check 
whether they are consistent with belonging to the Orphan stream.
\footnote{So far, the only existing PM models in the literature are those 
of \citet{new10}.} 
The \citet{new10} model was tailored to fit the observed positions, 
distances, and line-of-sight velocities of Orphan stream stars under 
a realistic MW potential, and therefore the PMs predicted by this model 
are expected to provide a reasonable first order guess of the true PMs.

\begin{figure*}
\epsscale{1.1}
\plottwo{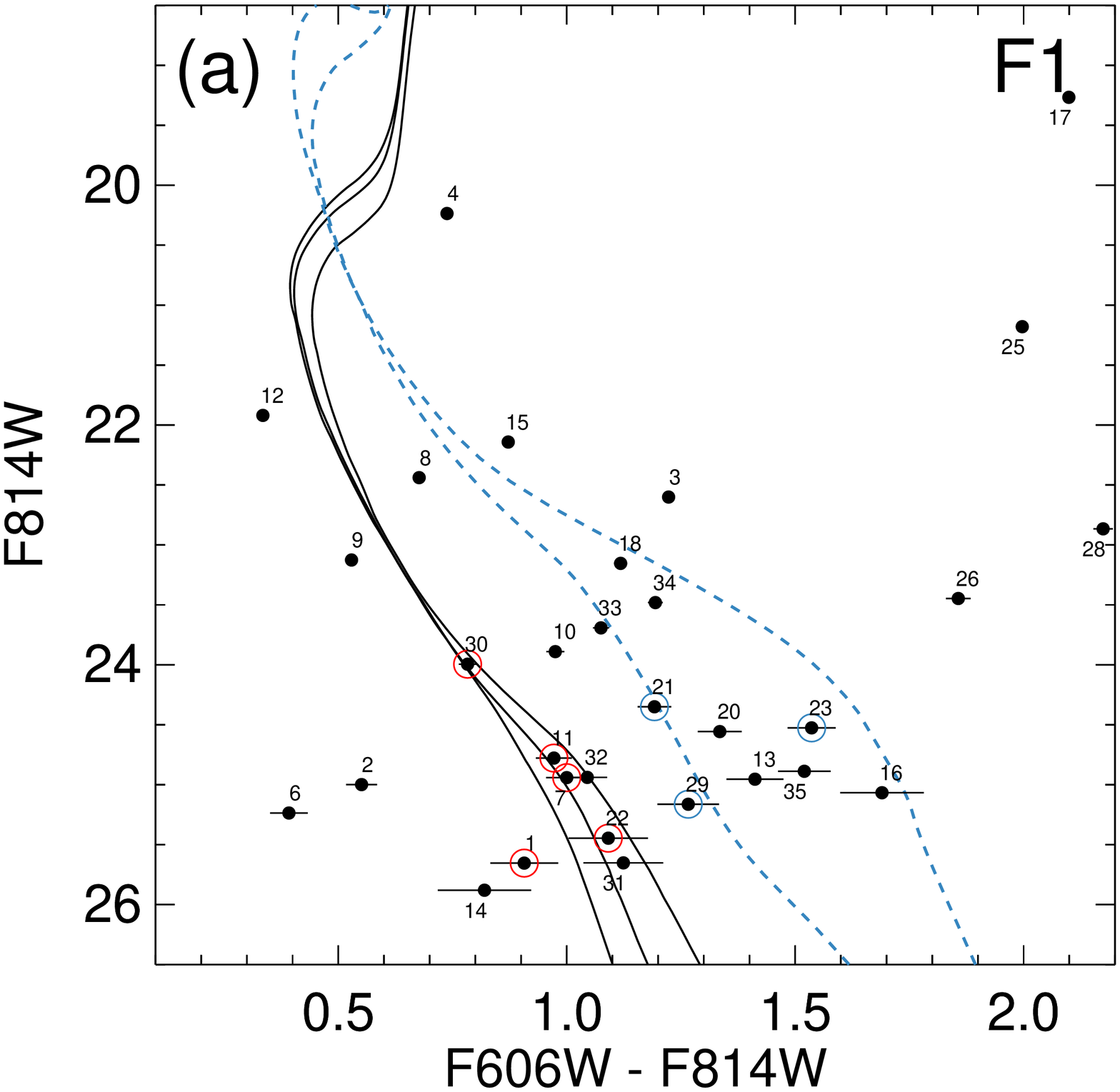}{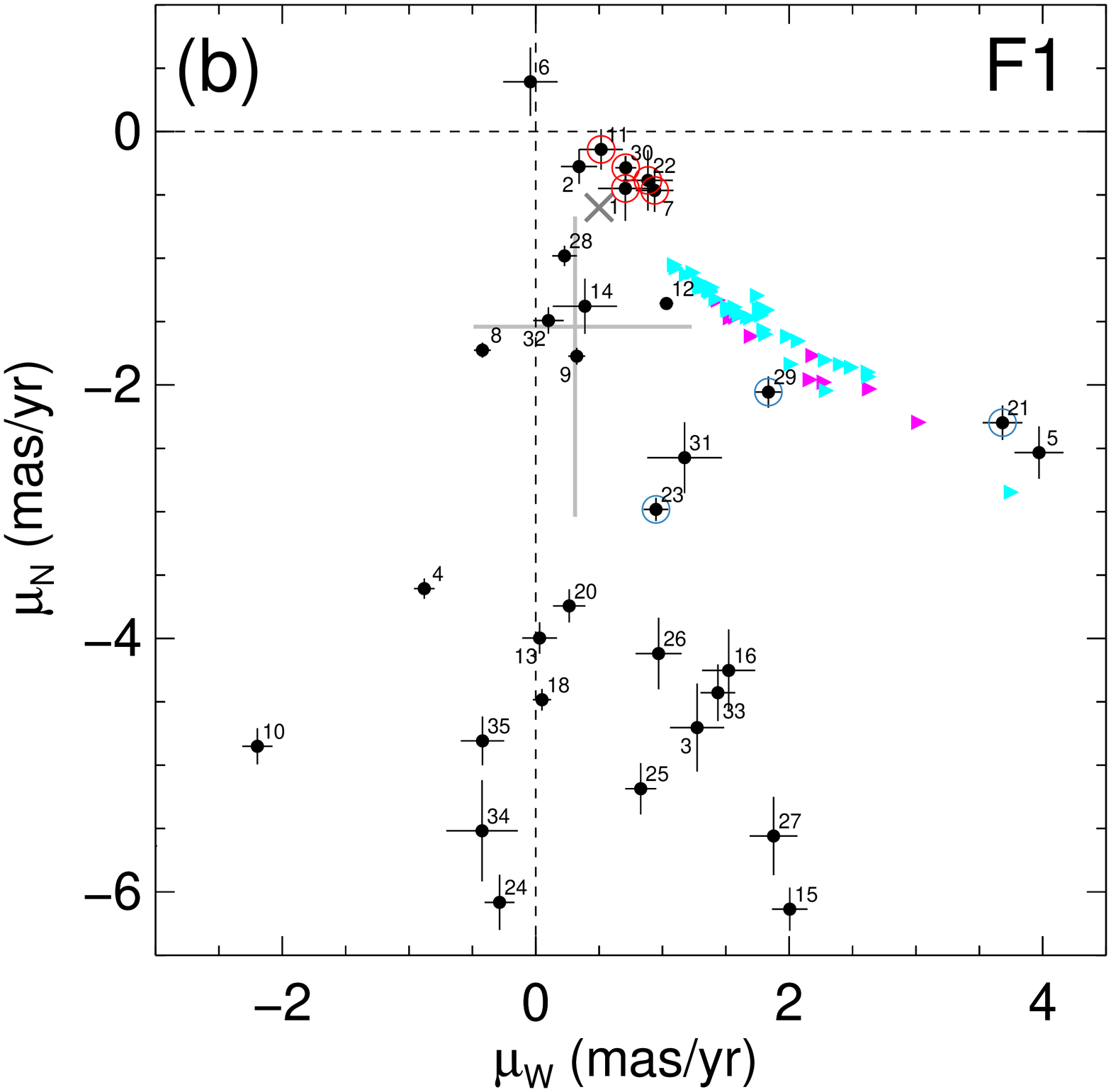}
\caption{Selection of stars associated with stellar streams in {\it
  ORPH-F1}. Panel~(a) shows the observed $I$ vs.~$V-I$ CMD and 
  panel~(b) shows the observed PM diagram. The observed stars in 
  the field are shown as black symbols in all panels, with the 
  associated error bars shown in both panels. Numbers are provided 
  for each star in each panel to aid cross identification. Not all 
  stars in panel~(a) are shown in panel~(b) due to their PMs being outside 
  the range of panel~(b). For most stars, the F814W magnitude errors 
  in panel~(a) are smaller than the plot symbols. In panel~(a), the 
  black solid curves are fiducial isochrones for old (12 Gyr) 
  populations with metallicities of [Fe/H] = $-1.5$, $-2.0$, and 
  $-2.5$ (from right to left) at a distance of 34 kpc. 
  The blue dashed curves are fiducial 
  isochrones for an old and metal-poor (11 Gyr, [Fe/H] $=-1.8$), 
  and an intermediate-age and metal-rich (5 Gyr, [Fe/H] $=-0.5$) 
  populations at distances of 17 kpc. The black solid curves are used 
  for identifying the Orphan stream population, while the blue dashed 
  curves are used for identifying potential Sgr Stream stars.  
  Red and blue circles in both panels highlight stars that are 
  identified as potential Orphan and Sgr Stream stars, respectively.
  The cross mark in panel~(b) indicates the predicted PM from 
  the orbital model by \citet{new10}. In the same panel, we also show 
  the \citet{law10} $N$-body model leading-arm particles in right-facing 
  triangles to provide PM predictions of the Sgr faint arm at this 
  field location assuming that both the bright and faint arms move in 
  parallel on the sky (see text for details). These particles are color 
  coded following \citet{law10} such that different colors represent the time 
  at which a given debris particle became unbound from the Sgr dSph: 
  magenta for 1.5--3 Gyr ago; and cyan for 3--5 Gyr ago. The PMs of the 
  model particles are offset from the original position such that their  
  distances are decreased by 3 kpc to account for the difference in 
  distance between the Sgr bright and faint arms. The gray cross shows 
  the 68 confidence intervals in each coordinate for MW halo star PMs 
  drawn from Besan\c{c}on models, chosen to meet our CMD selection 
  criteria for stars on the Orphan and Sgr streams. An expectation value 
  of 4 MW halo stars is predicted within the area spanned by the cross.
  \label{fig:field1}}
\end{figure*}

In addition to the stars of interest, our target fields will also 
contain MW stars in the foreground and background. It is therefore 
important to assess the number of MW stars in our PM diagrams. 
Foreground stars in the MW disk are in general not a 
concern since most of them are typically far redder than the isochrones 
used for selecting old stellar populations. Moreover, their PM 
distribution has a large spread of several $\masyr$ and do not clump 
in PM space \citep[see, e.g.][]{dea13}. Halo stars of the MW are more 
difficult to separate from stellar streams since they share similar 
properties in terms of stellar populations (and distances in some cases). 
Typically, there are a handful of MW halo stars predicted 
per field with a dispersion of order $\sim 1\masyr$ as we have shown 
in \citet{soh15}. To quantitatively predict the PM distribution of MW 
halo stars in each of our fields, we followed the same methodology used 
in that study. In short, we adopted the Besan\c{c}on model \citep{rob03} 
for each field location, selected model stars that satisfy our CMD 
selection criteria, and estimated the number of MW halo stars in each  
ACS/WFC field. In each of our PM diagrams (Figures~\ref{fig:field1}b, 
\ref{fig:field2}b, \ref{fig:field3}b, and \ref{fig:field4}b), we also 
show the median value and the 68\%\ confidence interval of the PM 
distribution of MW halo stars as a gray cross. A general discussion 
of the predicted PMs of MW stars in \hst\ fields is detailed in 
Section 3.1.3 of \citet{soh15}.

Finally, if we find a significant excess of stars against our estimates 
of both Orphan stream and MW halo stars, there is a possibility of 
other kinematical groups existing in our fields. In this case, we go  
back to the CMD to check for additional populations of stars at any given 
distance. For this, we use old and metal-poor isochrones because all 
stellar streams found in the MW halo so far are predominantly old and 
metal poor.

%

\subsection{{\it ORPH-F1}}
\label{sec:field1}

In Figure~\ref{fig:field1}, we plot CMDs and PM diagrams for the 
{\it ORPH-F1} field. The expected heliocentric distance to the Orphan 
Stream for this field location is 34 kpc and a reddening of 
$E(B-V) = 0.023$ was used. The isochrones representing an Orphan-like 
population are plotted as black curves in Figure~\ref{fig:field1}a. 
Among the stars consistent with these isochrones, we find five (marked 
in red circles) that are clustered around $(\muw, \mun) = (0.8, -0.5) 
\masyr$. These five stars are close to the PM prediction (cross symbol 
in Figure~\ref{fig:field1}b) for this field by \citet{new10}, and we 
identify them as being associated with the Orphan stream. 
We further analyze the resulting 2-d motion on the sky for these
stars in Section~\ref{sec:2dmotion}.

Upon inspecting the CMD and PM diagrams and further quantitatively 
comparing them to the predictions by the Besan\c{c}on model, we found 
that there is a general excess of stars. There are two ways of 
viewing this: an excess in the CMD or an excess in the PM diagram.
First, in the color range 
$1.0 < (F606W-F814W) < 1.7$ and magnitude range $23 < F814W < 25.5$,
the Besan\c{c}on model predicts only four stars in our field of view, 
but we have a count of 12 stars in that color and magnitude range 
in Figure~\ref{fig:field1}a. Second, we also find that while 
the Besan\c{c}on model predicts only one star in total 
(regardless of which MW component it belongs to) per ACS/WFC field 
in the the PM range of $\mun < -3.5 \masyr$, our PM diagram 
(Figure~\ref{fig:field1}b) shows 14 stars significantly in excess
of model predictions for the MW disk and a smooth halo.
We first discuss the excess as seen in the CMD, and then explore 
the excess in the PM diagram.

Our {\it ORPH-F1} field is located close to the ``faint arm'' of the 
Sgr Stream (see Figure~\ref{fig:fields}) on the sky, so we expect to 
find stars from this component. So far, there is no successful model 
that explains the bifurcation of the Sgr Stream, but the Sgr 
faint arm has been mapped in both the Northern \citep{bel06} and 
Southern hemispheres \citep{kop12,sla13,deb15}. The faint arm has been 
consistently measured to be about 5 kpc closer to us than the bright 
arm at any given Sgr longitude \citep[see e.g., Figure~7 of][]{sla13}. 
In the Sgr coordinates defined by \citet{maj03}, our {\it ORPH-F1} 
field is located at $\Lambda_{\odot} = 220\fdg7$, and the heliocentric 
distance of the bright arm at this Sgr longitude is $\sim 22$ kpc 
based on Figure~4 of \citet{bel14}. Therefore, stars on the Sgr faint 
arm, if present, are expected to be at a distance of $\sim 17$ kpc in 
this part of the sky. The Sgr bright arm is known to have a mixed 
population of stars: an old and metal-poor population $+$ an 
intermediate-age metal-rich population. In our Sgr stream study 
\citep{soh15}, we have found that stars from both of these populations 
are detected even in the small fields of \hst. We therefore overlaid  
isochrones of (age, [Fe/H]) $=$ (12 Gyr, $-1.5$) and (5 Gyr, $-0.5$) 
with distances of 17 kpc in our CMD, and first selected 10 stars that 
lie close to or in between these two isochrones. We then inspected 
the 2-d motions of these 10 stars on the sky and compared them with 
the Sgr faint arm to select candidate member stars.

%
\begin{figure}
\epsscale{1.15}
\plotone{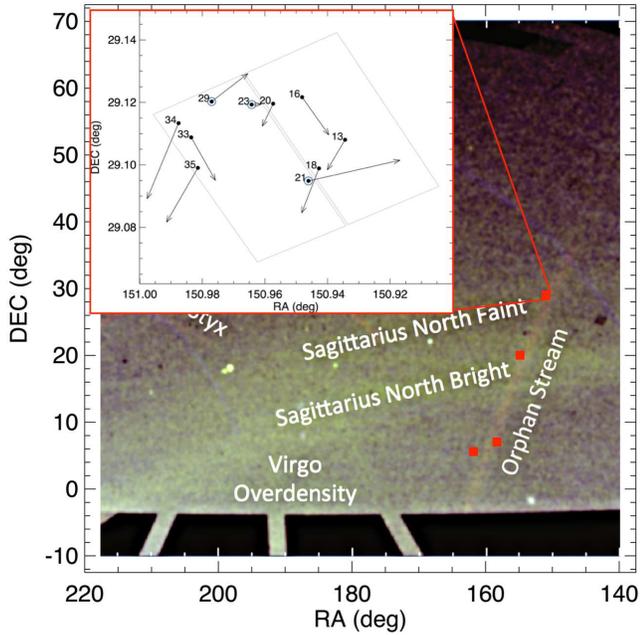}
\caption{Same as Figure~\ref{fig:fields}, but the inset now shows 
         the 2-d motion on the sky (arrows) of individual stars 
         selected via CMD as being consistent with a Sgr-like 
         population at the distance of 17 kpc. The gray lines 
         show the boundary of the ACS/WFC images on the sky, and 
         black dots are the CMD-selected stars as discussed in the 
         text. Each star is identified using the same number as 
         in Figure~\ref{fig:field1}.         
         The length of each vector is scaled in a relative sense 
         to show the difference in speed assuming that all stars are 
         at distances of 17 kpc. The stars with blue circles are 
         possible candidates of the Sgr faint arm based on their 
         based on their 2-d motions.
         \label{fig:SgrFaint}
        }
\end{figure}
%

While we currently have no existing information on the predicted 
PM of the Sgr faint arm, we can select stars based on the fact that
any star that belongs to the Sgr faint arm should follow the direction 
of the overall distribution of the stream (extending from southeast 
to northwest on the sky) to first order.
The observed PM of any star on the sky includes a contribution from 
the solar motion, so we need to subtract that to get the 2-d 
motion on the sky. For this, we assumed a distance to the Sun from 
the Galactic center of $R_{0} = 8.29 \kpc$, a circular velocity of 
the local standard of rest (LSR) of $V_0 = 239 \kms$ \citep{mcm11}, 
and a solar peculiar velocity with respect to the LSR of 
$(U, V, W)_{\rm pec} = 11.10, 12.24, 7.25) \kms$ \citep{sch10}.
The PM contribution of the solar motion also depends on the distance 
to the target object, so we used the distance of 17 kpc.
At this distance, the PM contribution from the solar motion toward 
the {\it ORPH-F1} field is $(\mu_{\rm W}, \mu_{\rm N}) = 
(0.63, -2.99)~\masyr$. This motion was {\it subtracted} from the 
observed PM of each star to obtain a net 2-d motion on the sky.
The resulting 2-d vectors compared to the Sgr faint arm are 
shown in Figure~\ref{fig:SgrFaint}. We find that the three stars 
circled in blue in Figure~\ref{fig:field1} (stars 21, 23, and 29)
have 2-d motions consistent with the Sgr faint arm. However, we note 
that the PMs of these three stars are not clumped together in 
Figure~\ref{fig:field1}b which indicates it is unlikely that all three 
stars are members of the Sgr faint arm. 
Without a detailed dynamical model that explains the Sgr faint arm, 
it is difficult to determine which of these three stars belong to 
the Sgr stream. We can however obtain a first-order guess of where the 
stars should be placed in the PM diagram assuming that both the bright 
and faint arms of the Sgr move in parallel on the sky. For this, 
we selected model particles from the $N$-body model of \citet{law10} 
that are located within 1$\degr$ from the same Sgr longitude as our 
{\it ORPH-F1} field, and corrected the model PM predictions for 
the difference in distance (3 kpc) between the bright and faint arms.
The resulting model particles are shown in Figure~\ref{fig:field1}b. 
Based on comparing the model and observations, star 29 is most likely 
a star that belongs to the Sgr faint arm, while the other two stars 
are less likely to be so. 

The excess of stars in our PM diagram for $\mun < -3.5 \masyr$ 
indicates that there may be substructures (stellar streams or 
kinematically-cold groups) other than the Sgr and Orphan Streams in 
the halo in this part of the sky at distances closer than 17 kpc. 
However, we have not found any obvious PM clumps in this field 
that correspond to the same stellar population at the same distance 
based on the CMD.

\subsection{{\it ORPH-F2}}
\label{sec:field2}

%
\begin{figure*}
\epsscale{1.1}
\plottwo{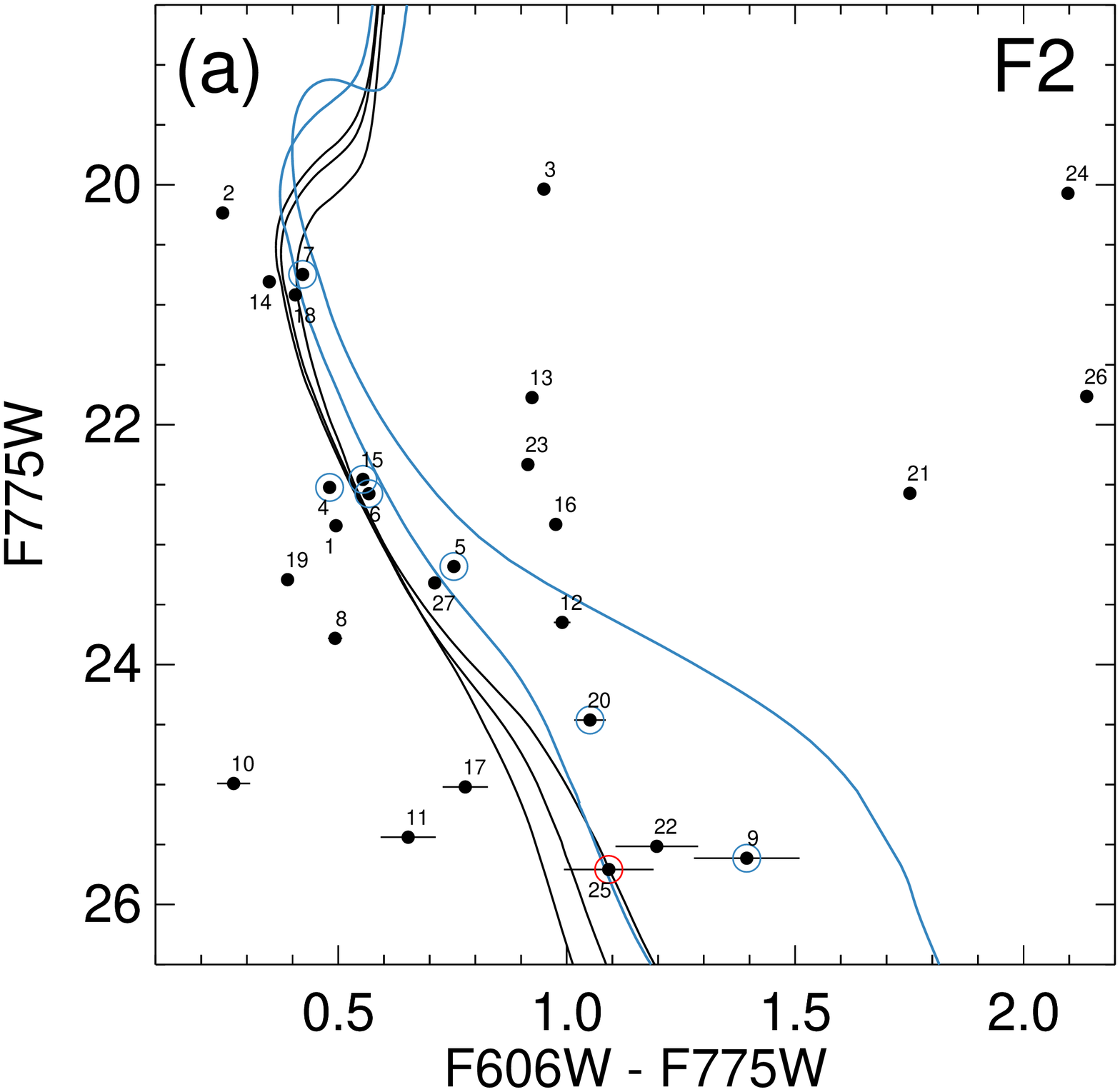}{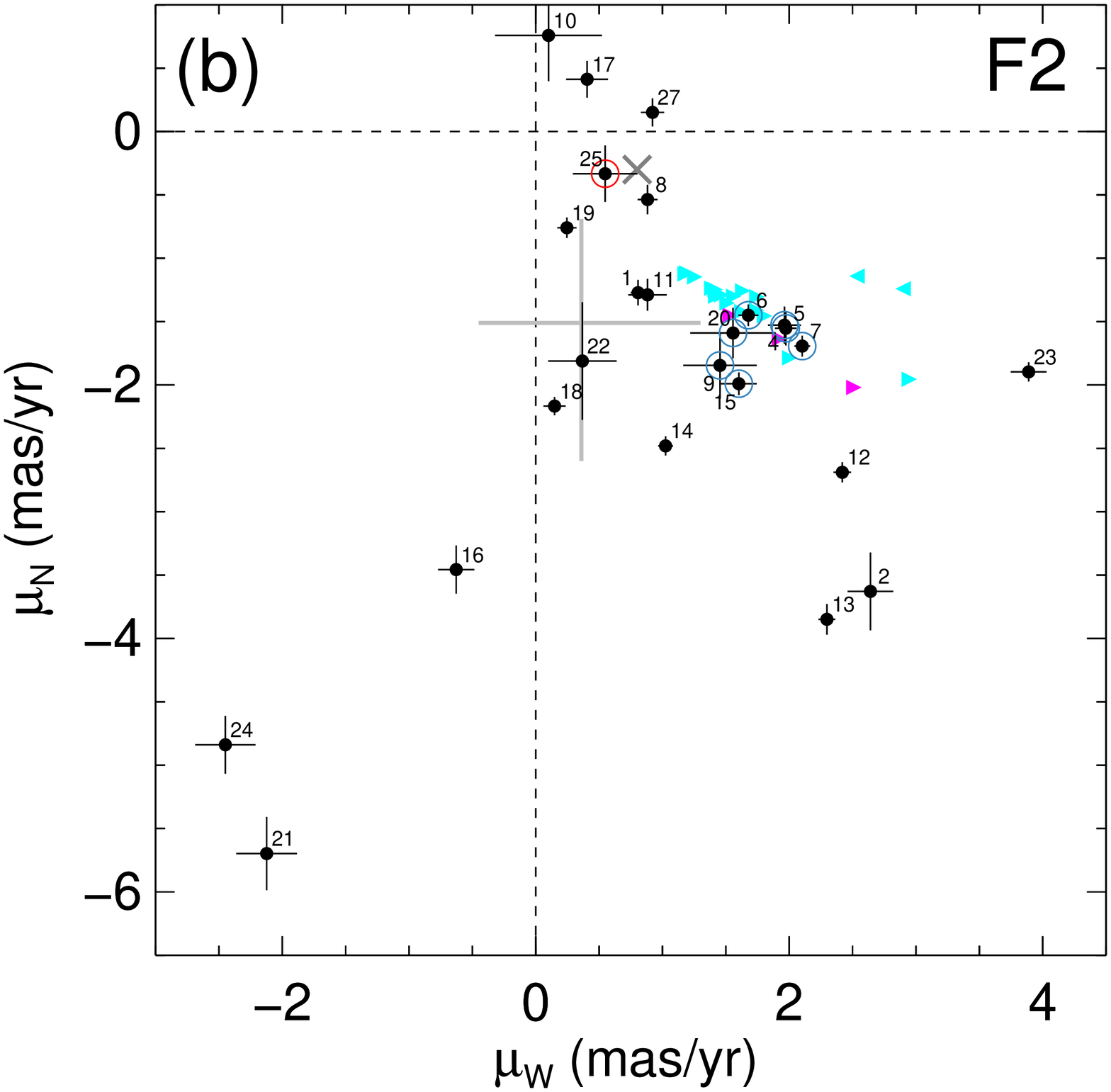}
\caption{Selection of stars associated with stellar streams in the 
  {\it ORPH-F2} field. The panels and symbols are similar to those in
  Figure~\ref{fig:field1}. 
  The black solid curves in panel~(a) are the same isochrones as 
  used in Figure~\ref{fig:field1} but for a distance of 29 kpc.
  For the Sgr Stream, solid blue lines are used in panel~(a) 
  instead of dashed blue lines to represent the Sgr bright arm, 
  and the distance of 23 kpc was adopted for these isochrones.
  In panel~(b), we also show the \citet{law10} $N$-body model particles in 
  right- (leading) or left-facing (trailing) triangles. These particles 
  are color coded following \citet{law10} such that different colors 
  represent the time at which a given debris particle became unbound from 
  the Sgr dSph: magenta for 1.5--3 Gyr ago; and cyan for 3--5 Gyr ago.  
  The gray cross is the same as in Figure~\ref{fig:field1}.
  An expectation value of 3 MW halo stars is predicted within the area 
  spanned by the gray cross.
  \label{fig:field2}}
\end{figure*}

Figure~\ref{fig:field2} shows the CMD and PM diagram for the 
{\it ORPH-F2} field. For the CMD, we adopted the same fiducial isochrone 
as used for the {\it ORPH-F1} field but with a distance of 29 kpc. 
Unlike the case for {\it ORPH-F1} field, we do not find any noticeable 
clump in the PM diagram. There is only one star (circled in red) that is 
consistent with being an Orphan-like 
population, and close enough to the PM prediction by \citet{new10}.
Without further information, it is difficult to judge whether this 
star actually belongs to the Orphan stream or is just a random halo star. 
Therefore, the identification of Orphan stream stars in this field 
is tentative.

The {\it ORPH-F2} field is located on the main Sgr stream (or the 
bright arm) at $\Lambda_{\odot} = 226\fdg4$, and we expect Sgr Stream 
stars to be present in this field. 
We have overlaid the same age+metallicity isochrones used in the 
{\it ORPH-F1} field for identifying Sgr stream stars, but at a distance 
of 22 kpc based on Figure~4 of \citet{bel14}. Among the stars consistent 
with these isochrones, seven stars are easily identified as a conspicuous 
clump in the PM diagram. These are plotted in blue circles in 
Figure~\ref{fig:field2}. For comparison, we also plot the \citet{law10} 
$N$-body model particles that are within 1$\fdg$5 in both right 
ascension and declination from the center of our HST field. 
The observed PM clump is very close to the model particles as expected,
and similar to the findings for Sgr {\it FIELD 4} in \citet{soh15} 
at $\Lambda_{\odot} = 251\fdg1$. Beyond these, no other significant 
excess has been found in the PM diagram.


\subsection{{\it ORPH-F3}}
\label{sec:field3}

%
\begin{figure*}
\epsscale{1.1}
\plottwo{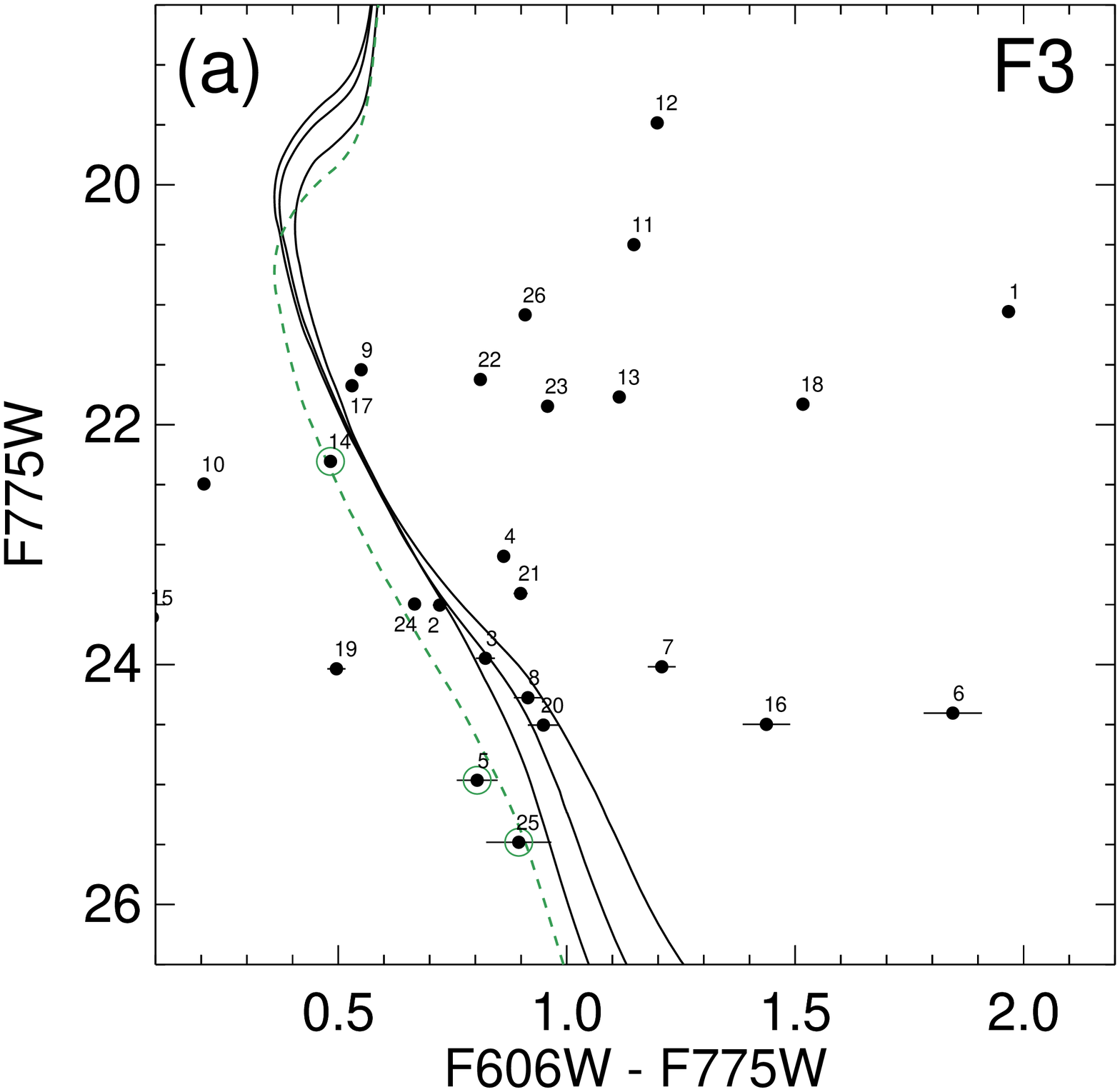}{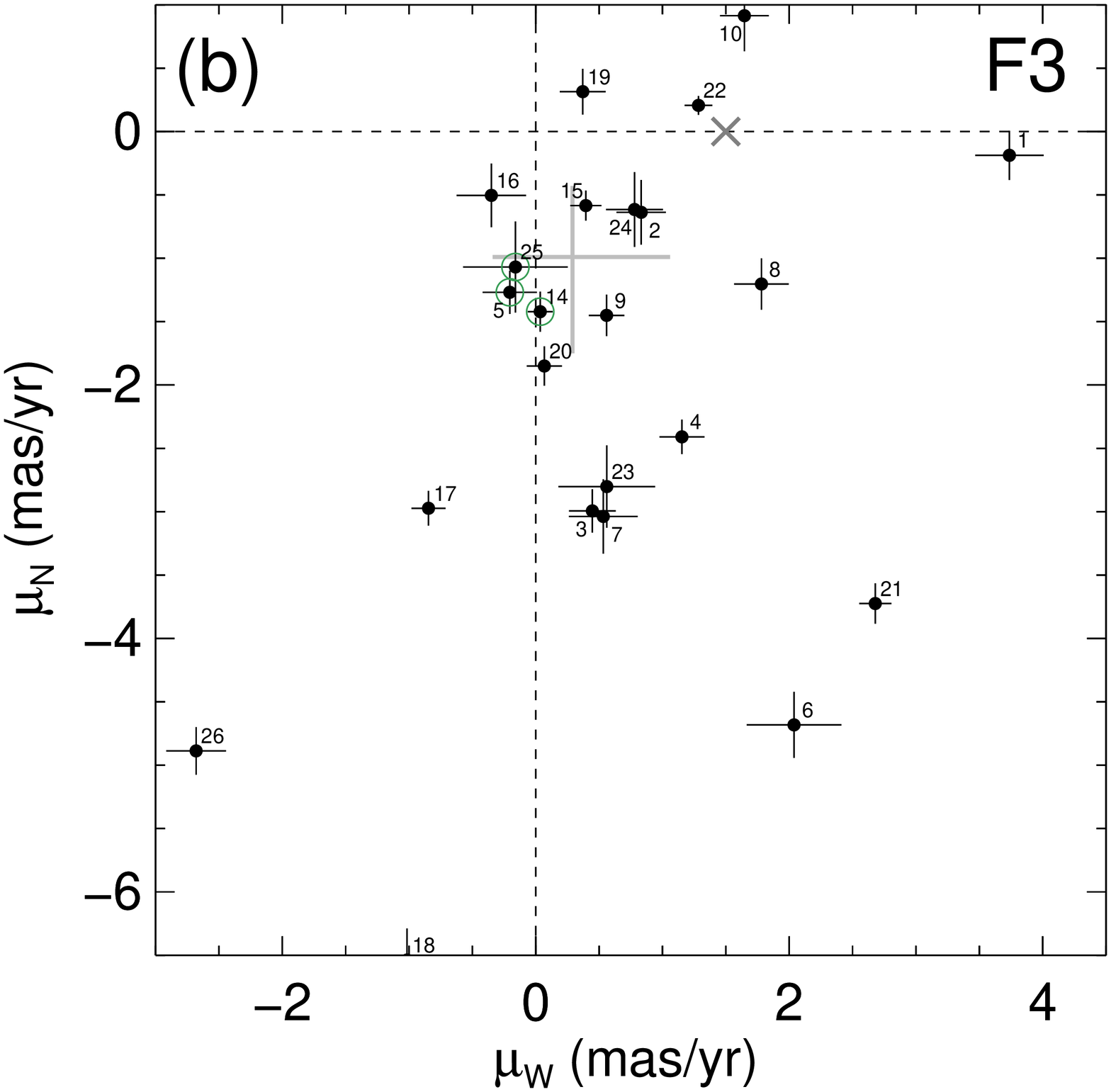}
\caption{Selection of stars associated with stellar streams in the 
  {\it ORPH-F3} field. The panels and symbols are similar to those in
  Figure~\ref{fig:field1}. The black solid curves in panel~(a) are the 
  same isochrones as used in Figure~\ref{fig:field1} but for a distance 
  of 24 kpc. The dashed green curve is a fiducial isochrone for 
  a population with an age of 12 Gyr and a metallicity of [Fe/H] 
  $= -2.5$ at a distance of 32 kpc. This isochrone is used to identify 
  stars possibly associated with the ``Parallel'' stream (green circles).
  The gray cross is the same as in Figure~\ref{fig:field1}, but for 
  CMD selection criteria of the newly identified stream. An expectation 
  value of 2 MW halo stars is predicted within the area spanned by the 
  gray cross.
  \label{fig:field3}}
\end{figure*}

Figure~\ref{fig:field3} shows the CMD and PM diagram for the 
{\it ORPH-F3} field. If present, Orphan stream stars should be at 
a distance of 24 kpc. We used the fiducial isochrones to identify 
Orphan stream stars, and found four stars consistent with the isochrones 
within their photometric errors. However, all of these stars are 
far away from the PM prediction of \citet{new10}. There is 
only one star (22) that is close enough to the model prediction 
to be considered a member of the Orphan stream, but its color 
($F606W-F775W \simeq 0.8$) is too red for its brightness to be 
considered an Orphan stream star. We conclude that there is no star 
associated with the Orphan stream in this field. Given that this field 
is about 1$\fdg$2 away in the orthogonal direction from the densest 
part of the Orphan stream, and given that the Orphan stream is a thin 
stellar stream, it is unsurprising to find no star in a small \hst\ 
ACS field.

Upon inspecting the CMD and PM diagram further, we found stars that 
possibly belong to a kinematically cold component at the distance of 
$\sim 32$ kpc from us. The three stars, shown in green circles, are 
well described by an isochrone with (age, [Fe/H], distance) 
$=$ (12 Gyr, $-2.5$, 32 kpc), and form a clump in the PM 
diagram at $(\mu_{\rm N}, \mu_{\rm W}) \approx (-0.1, -1.3)$. 
We discuss the nature of this clump in Section~\ref{sec:2dmotion}.
We further attempted to identify other PM clumps in this field, 
but did not find any that seem to belong to the same stellar population  
based on shifting old and metal-poor isochrones in the CMD.

\subsection{{\it ORPH-F4}}
\label{sec:field4}

%
\begin{figure*}
\epsscale{1.1}
\plottwo{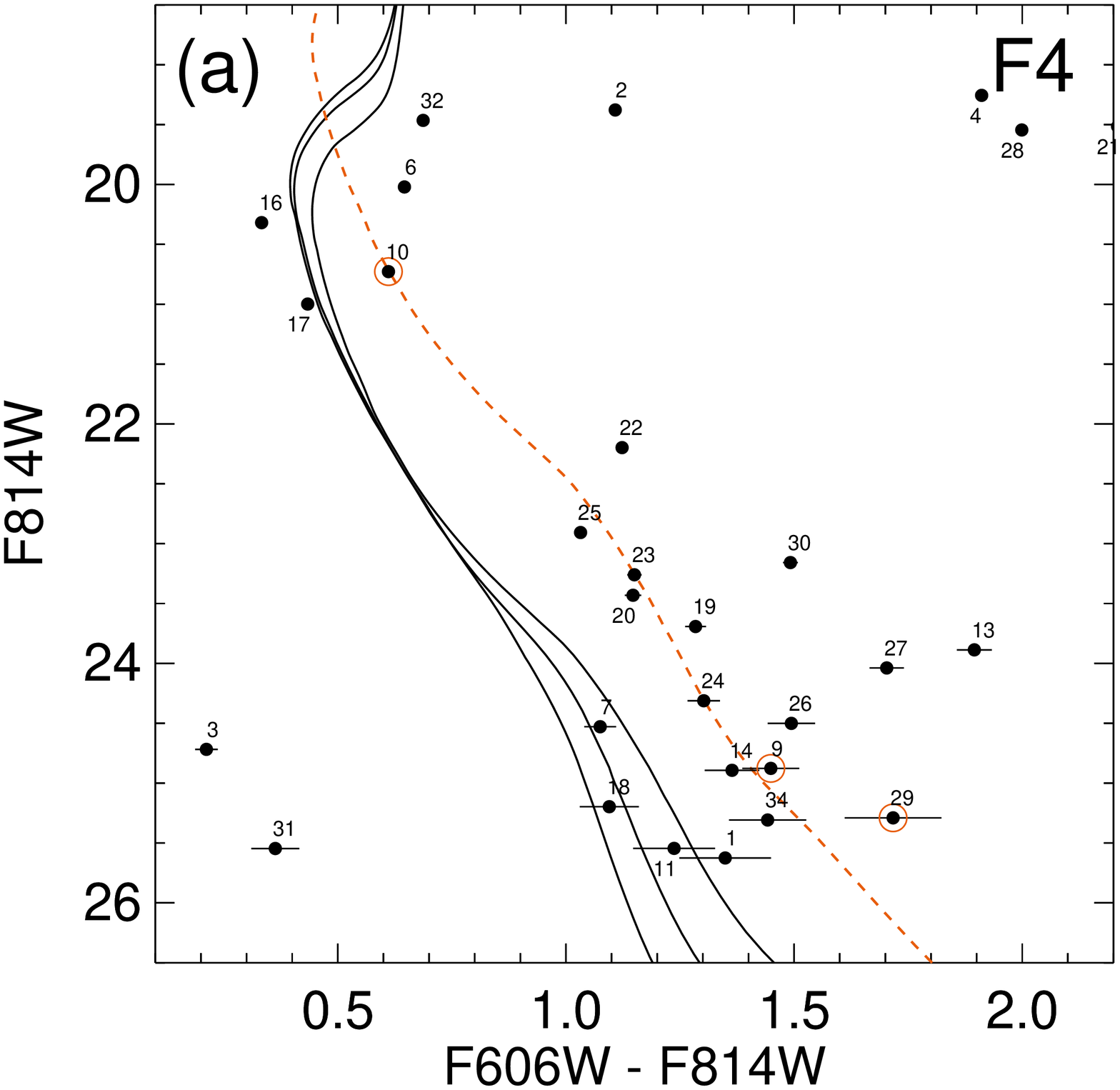}{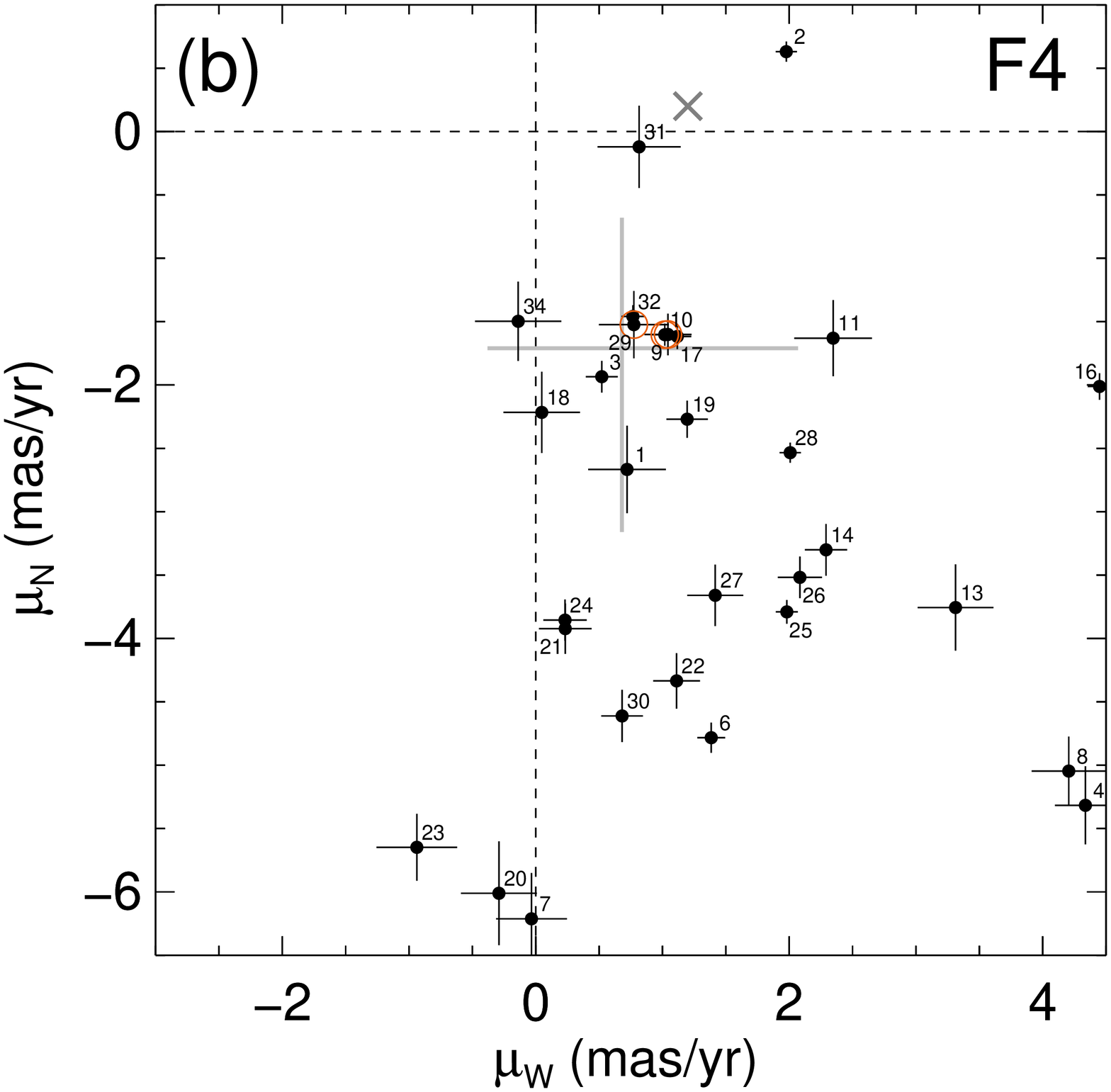}
\caption{Selection of stars associated with stellar streams in the 
  {\it ORPH-F4} field. The panels and symbols are similar to those in
  Figure~\ref{fig:field1}.
  The black solid curves in panel~(a) are the 
  same isochrones as used in Figure~\ref{fig:field1} but for a distance 
  of 23 kpc. The dashed orange curve is a fiducial isochrone for 
  a population with an age of 12 Gyr and a metallicity of [Fe/H] 
  $= -1.5$ at a distance of 12 kpc. This isochrone is used to identify 
  stars possibly associated with the Lethe Stream (orange circles).
  The gray cross is the same as in Figure~\ref{fig:field1}, but for 
  CMD selection criteria of the Lethe Stream. An expectation value of 
  4 MW halo stars is predicted within the area spanned by the gray cross.
  \label{fig:field4}}
\end{figure*}

Figure~\ref{fig:field4} shows the CMD and PM diagram for the 
{\it ORPH-F4} field. This field is located almost on the opposite side 
of the {\it ORPH-F3} field with respect to the Orphan stream, and is 
about 1$\fdg$6 away from the densest part of the Orphan stream. 
Therefore, the expected number of Orphan stream stars is even lower 
than in the case of {\it ORPH-F3}. The only star (\#31) close to 
the model prediction in the PM diagram of Figure~\ref{fig:field4}b 
is too blue in color for its brightness to be a Orphan 
stream star, so we conclude that we do not detect any star that 
belongs to the Orphan stream in this field.

As with the {\it ORPH-F3} field, we searched for additional kinematically 
cold components, and found 3 stars that possibly belong to one at the 
distance of $\sim 12$ kpc.\footnote{Note that we used the brightest star 
(14) among the three to estimate the distance to this kinematically cold 
component because that star has the smallest color and magnitude error.} 
The 3 stars circled in orange in Figure~\ref{fig:field4}a are well described 
by an isochrone with (age, [Fe/H], distance) $=$ (12 Gyr, $-1.5$, 11 kpc), 
and form a clump in the PM diagram at $(\mu_{\rm N}, \mu_{\rm W}) \approx 
(1.0, -1.7)$. We discuss the nature of this clump in Section~\ref{sec:2dmotion}.

We also found a significant 
excess of stars in our PM diagram (Figure~\ref{fig:field4}b) at 
$\mun < -3 \masyr$. Our diagram shows 15 stars, but the Besan\c{c}on 
model predicts only two stars in total per ACS/WFC field in the same 
PM range. However, we were not able to further identify PM clumps 
that belong to the same stellar population at the same distance.

\section{Average Motions of Stars in Our Target Fields}
\label{sec:2dmotion}

%
\begin{deluxetable*}{rccccccc}
\tablecaption{Proper motions and photometry of stars identified as 
stream members in each field\label{tab:pmtbl}}
\tablehead{
 \colhead{}    & \colhead{R.A. (J2000.0)} & \colhead{Decl. (J2000.0)} & \colhead{$\muw$}     & \colhead{$\mun$}      & \colhead{F814W or F775W\tablenotemark{a}}  & \colhead{F606W}     & \colhead{} \\
 \colhead{No.} & \colhead{(hh:mm:ss.sss)} & \colhead{(dd:mm:ss.ss)}   & \colhead{($\masyr$)} & \colhead{($\masyr$)} & \colhead{(VEGAMAG)}                        & \colhead{(VEGAMAG)} & \colhead{Stream}
   }
\startdata
\cutinhead{{\it ORPH-F1}}
   1 & 10:03:37.786 & $+$29:05:45.97 & \phm{$-$}0.71 $\pm$ 0.22 &  $-$0.45 $\pm$ 0.26 & 25.65 $\pm$ 0.03 & 26.56 $\pm$ 0.07 & Orphan \\
   7 & 10:03:46.959 & $+$29:08:12.53 & \phm{$-$}0.94 $\pm$ 0.15 &  $-$0.46 $\pm$ 0.17 & 24.94 $\pm$ 0.02 & 25.94 $\pm$ 0.04 & Orphan \\
  11 & 10:03:43.184 & $+$29:06:07.52 & \phm{$-$}0.52 $\pm$ 0.17 &  $-$0.14 $\pm$ 0.16 & 24.78 $\pm$ 0.02 & 25.75 $\pm$ 0.04 & Orphan \\
  22 & 10:03:45.787 & $+$29:05:00.63 & \phm{$-$}0.88 $\pm$ 0.20 &  $-$0.39 $\pm$ 0.24 & 25.45 $\pm$ 0.04 & 26.54 $\pm$ 0.08 & Orphan \\
  30 & 10:03:54.727 & $+$29:06:57.56 & \phm{$-$}0.71 $\pm$ 0.08 &  $-$0.29 $\pm$ 0.09 & 24.00 $\pm$ 0.01 & 24.78 $\pm$ 0.02 & Orphan \\
  29 & 10:03:54.485 & $+$29:07:12.94 & \phm{$-$}1.84 $\pm$ 0.10 &  $-$2.06 $\pm$ 0.12 & 25.16 $\pm$ 0.02 & 26.43 $\pm$ 0.06 & Sgr Faint? \\
\cutinhead{{\it ORPH-F2}}
   4 & 10:19:07.628 & $+$20:01:17.32 & \phm{$-$}1.96 $\pm$ 0.13 &  $-$1.53 $\pm$ 0.15 & 22.52 $\pm$ 0.01 & 23.00 $\pm$ 0.01 & Sgr \\
   5 & 10:19:08.159 & $+$20:01:29.91 & \phm{$-$}1.97 $\pm$ 0.09 &  $-$1.55 $\pm$ 0.14 & 23.18 $\pm$ 0.01 & 23.93 $\pm$ 0.01 & Sgr \\
   6 & 10:19:09.386 & $+$20:02:05.84 & \phm{$-$}1.68 $\pm$ 0.08 &  $-$1.45 $\pm$ 0.09 & 22.57 $\pm$ 0.01 & 23.14 $\pm$ 0.01 & Sgr \\
   7 & 10:19:08.624 & $+$20:01:17.95 & \phm{$-$}2.10 $\pm$ 0.06 &  $-$1.69 $\pm$ 0.08 & 20.75 $\pm$ 0.00 & 21.17 $\pm$ 0.00 & Sgr \\
   9 & 10:19:12.261 & $+$20:03:44.42 & \phm{$-$}1.45 $\pm$ 0.29 &  $-$1.85 $\pm$ 0.26 & 25.61 $\pm$ 0.04 & 27.01 $\pm$ 0.11 & Sgr \\
  15 & 10:19:12.869 & $+$20:01:01.52 & \phm{$-$}1.60 $\pm$ 0.14 &  $-$1.99 $\pm$ 0.09 & 22.46 $\pm$ 0.01 & 23.01 $\pm$ 0.01 & Sgr \\
  20 & 10:19:17.706 & $+$20:01:18.79 & \phm{$-$}1.56 $\pm$ 0.34 &  $-$1.59 $\pm$ 0.20 & 24.46 $\pm$ 0.02 & 25.51 $\pm$ 0.03 & Sgr \\
  25 & 10:19:19.969 & $+$20:01:04.59 & \phm{$-$}0.55 $\pm$ 0.26 &  $-$0.33 $\pm$ 0.22 & 25.71 $\pm$ 0.05 & 26.80 $\pm$ 0.09 & Orphan? \\
\cutinhead{{\it ORPH-F3}}
   5 & 10:33:17.340 & $+$07:01:29.63 &       $-$0.20 $\pm$ 0.21 &  $-$1.27 $\pm$ 0.17 & 24.96 $\pm$ 0.03 & 25.77 $\pm$ 0.04 & Parallel \\
  14 & 10:33:13.766 & $+$07:02:39.48 & \phm{$-$}0.04 $\pm$ 0.10 &  $-$1.42 $\pm$ 0.16 & 22.31 $\pm$ 0.00 & 22.79 $\pm$ 0.01 & Parallel \\
  25 & 10:33:08.859 & $+$07:02:43.71 &       $-$0.16 $\pm$ 0.41 &  $-$1.07 $\pm$ 0.36 & 25.48 $\pm$ 0.04 & 26.38 $\pm$ 0.06 & Parallel \\
\cutinhead{{\it ORPH-F4}}
   9 & 10:47:22.903 & $+$05:37:31.78 & \phm{$-$}1.04 $\pm$ 0.19 &  $-$1.60 $\pm$ 0.16 & 24.88 $\pm$ 0.02 & 26.33 $\pm$ 0.06 & Lethe \\
  10 & 10:47:19.105 & $+$05:35:57.16 & \phm{$-$}1.02 $\pm$ 0.10 &  $-$1.60 $\pm$ 0.09 & 20.73 $\pm$ 0.00 & 21.34 $\pm$ 0.00 & Lethe \\
  29 & 10:47:25.481 & $+$05:34:21.22 & \phm{$-$}0.77 $\pm$ 0.28 &  $-$1.52 $\pm$ 0.27 & 25.29 $\pm$ 0.03 & 27.01 $\pm$ 0.10 & Lethe
\enddata
\tablenotetext{a}{F814W for fields {\it ORPH-F1} and {\it ORPH-F4}, and F775W for fields {\it ORPH-F2} and {\it ORPH-F3}} 
\end{deluxetable*}
%

%
\begin{deluxetable}{lccccc}
\tablecolumns{6}
\tablecaption{Proper motion averages and dispersions\label{tab:pmavgs}}
\tablehead{
 \colhead{}       & \colhead{$\muw$}     & \colhead{$\mun$}     & \colhead{}                             & \multicolumn{2}{c}{$\sigma_{\rm 1-d}$\tablenotemark{b}} \\
 \cline{5-6}
 \colhead{Sample} & \colhead{($\masyr$)} & \colhead{($\masyr$)} & \colhead{$N_{\star}$\tablenotemark{a}} & \colhead{($\masyr$)} & \colhead{($\kms$)}
}
\startdata
\sidehead{{\it ORPH-F1}}
Orphan  &    0.74 $\pm$ 0.06 & $-$0.31 $\pm$ 0.05 & 5 & 0.14 &    23.0 \\
\hline
\sidehead{{\it ORPH-F2}}
Orphan\tablenotemark{c} & 0.55 $\pm$ 0.26 & $-$0.33 $\pm$ 0.22 & 1 & \nodata & \nodata \\
Sgr     &    1.91 $\pm$ 0.10 & $-$1.67 $\pm$ 0.09 & 7 & 0.17 &    16.4 \\
\hline
\sidehead{{\it ORPH-F3}}
Parallel & $-$0.01 $\pm$ 0.09 & $-$1.32 $\pm$ 0.08 & 3 & 0.14 &    20.9 \\
\hline
\sidehead{{\it ORPH-F4}}
Lethe  &    1.00 $\pm$ 0.05 & $-$1.60 $\pm$ 0.02 & 3 & 0.10 & \phn5.6 
\enddata
\tablenotetext{a}{Number of stream stars included in the PM calculations.}
\tablenotetext{b}{Average one-dimensional velocity dispersion, obtained from 
the combined West and North measurements, corrected for observational
scatter by subtracting the median random PM error bar in quadrature. The 
transformations from $\masyr$ to $\kms$ are based on the same distances as 
used for the isochrones in Figures~\ref{fig:field1}a--\ref{fig:field4}a.}
\tablenotetext{c}{Tentative detection of a single star.}
\end{deluxetable}

%
\begin{figure}
\epsscale{1.15}
\plotone{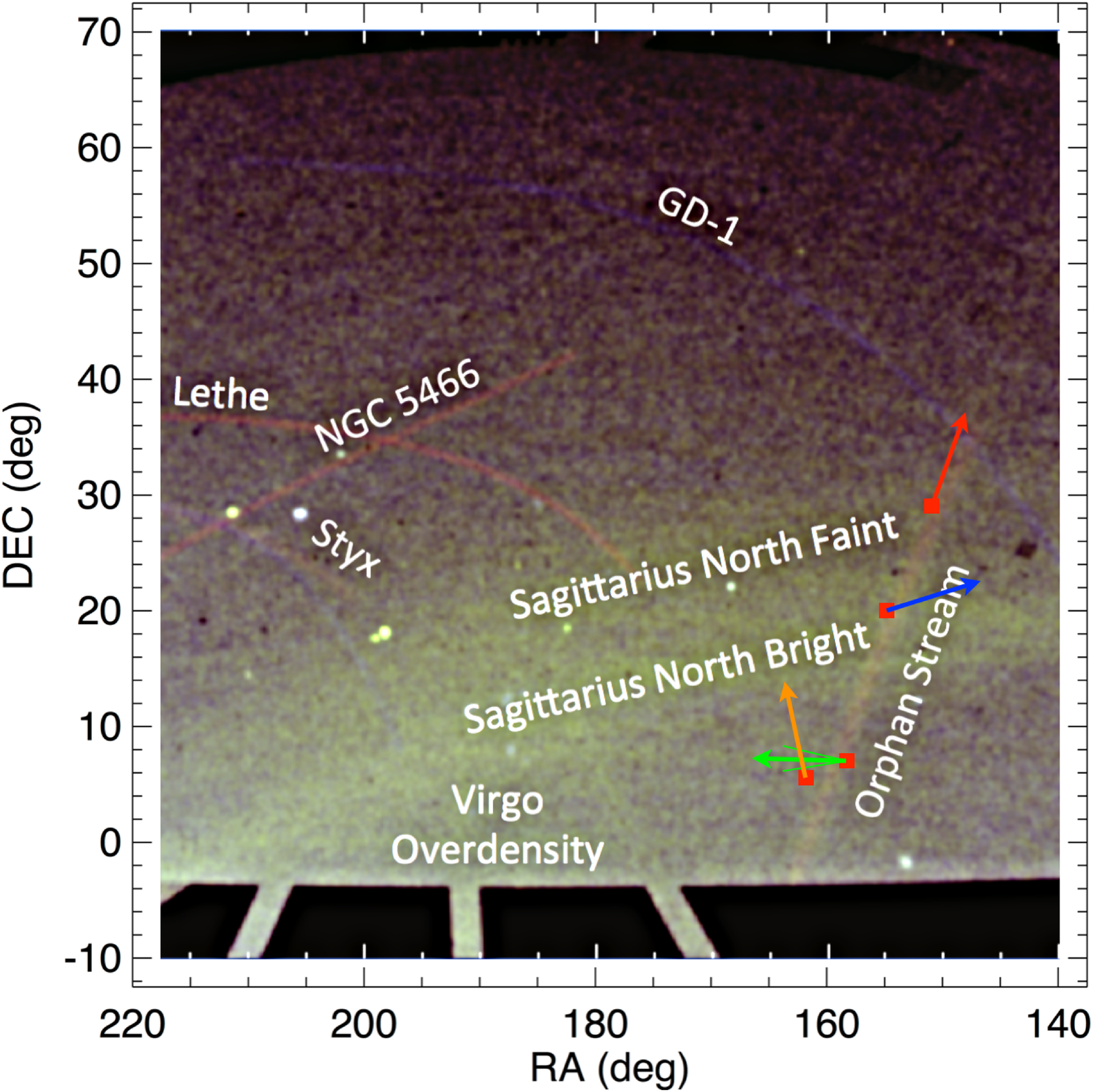}
\caption{Same as Figure~\ref{fig:fields}, but the directions of
         2-d motions of stars that (possibly) belong to different 
         streams are shown in colored arrows. Red and blue arrows are 
         for the Orphan and Sgr Streams, respectively.
         Green and orange arrows are for the kinematical components 
         found in fields {\it ORPH-F3} (the Parallel stream, a possible 
         new stream) and {\it -F4} (the Lethe stream), respectively.
         The magnitude of each vector was arbitrarily chosen to 
         show the direction of motion clearly, and does not represent 
         the actual space velocity of each component. The thin lines 
         next to the green arrow for field {\it ORPH-F3} represent the 
         1$\sigma$ error of the 2-d motions based on the PM uncertainties.
         For the other fields, similar 1$\sigma$-error lines overlap 
         with the arrows due to their small PM errors.
         \label{fig:fields_pmarrow}
        }
\end{figure}
%

The analysis in Section~\ref{sec:identification} has yielded a list of 
stars that belong to several different kinematical components in our 
target fields. In Table~\ref{tab:pmtbl}, we provide a list of 
individual proper motions and photometry for stars identified as 
members of various stellar streams based on our analysis in 
Section~\ref{sec:identification}.
Using this list, we first calculated the average PM of 
stars in each component and its associated uncertainty. When calculating 
average PMs for an individual sample, we adopted the error-weighted mean, 
and the error was computed by propagating the individual PM uncertainties. 
The results are presented in Table~\ref{tab:pmavgs}. 
For each component, we also list an estimate of the average intrinsic 
one-dimensional dispersion $\sigma$ transverse to the line-of-sight  
in both $\masyr$ and $\kms$. 

Using the average PM of each kinematical component, we can describe  
their 2-d motions on the sky. This allows us to visually confirm whether  
our measured PM is consistent with the directions of the spatial extents   
of the Orphan and Sgr Streams, or to associate the newly found 
kinematical components to other known streams. 
The observed PM of any star on the sky includes a contribution from 
the solar motion, so we need to subtract that to get the 2-d 
motion on the sky as we have demonstrated for individual stars in 
Section~\ref{sec:field1}. 
The PM contribution of the solar motion also depends on the distance 
to the target object, so we used the distance to each kinematical 
component as estimated in Section~\ref{sec:identification}.

\subsection{Orphan and Sagittarius Stream Stars}

The Orphan stream stars we found in the {\it ORPH-F1} field are at 
a distance of $\sim 34$ kpc, and the PM contribution from the solar 
motion at this distance is $(\mu_{\rm W}, \mu_{\rm N}) = 
(0.31, -1.50)~\masyr$. Therefore, the net 2-d motion of Orphan 
stream stars becomes $(\mu_{\rm W}, \mu_{\rm N}) = 
(0.43, 1.19)~\masyr$. The resulting vector of motion is 
illustrated in Figure~\ref{fig:fields_pmarrow} as the red arrow.
The direction of motion is well aligned with the Orphan Stream, 
and the Orphan stream stars found in the {\it ORPH-F1} 
field are consistent with being in the leading arm of the stream
in view of the discovery of the Orphan progenitor in the 
southern sky \citep{gri15}.

The Sgr Stream stars found in the {\it ORPH-F2} field are estimated 
to be at a distance of $\sim 23$ kpc, and the PM contribution from the 
solar motion at this distance is $(\mu_{\rm W}, \mu_{\rm N}) = 
(0.56, -2.10)~\masyr$. The net 2-d motion of the Sgr stream 
stars then becomes $(\mu_{\rm W}, \mu_{\rm N}) = 
(1.35, 0.43)~\masyr$, and this vector of motion is illustrated 
in Figure~\ref{fig:fields_pmarrow} as the blue arrow.
As with the Orphan stream case, the direction of motion is 
reasonably well aligned with the Sgr bright arm, and is 
consistent with motion along the leading arm.

\subsection{Field {\it ORPH-F3}: Possible Association with a 
Newly Identified Stellar Stream}

In the {\it ORPH-F3} field, we found three stars at the distance of 
$\sim 32$ kpc that likely belong to a group of stars moving along 
the same direction, given the tight clustering found in the PM 
diagram. To test the possibility of these three stars belonging 
to any known stellar stream, we first explore the 2-d motion on 
the sky. The PM contribution from the solar motion toward the 
{\it ORPH-F3} field at the distance of 32 kpc is $(\mu_{\rm W}, 
\mu_{\rm N}) = (0.47, -1.34)~\masyr$, so the resulting net 2-d 
motion of the three stars becomes $(\mu_{\rm W}, \mu_{\rm N}) = 
(-0.48, 0.02)~\masyr$. This implies a motion almost entirely to 
the east on the sky with only a slight motion toward the north. 
This vector of motion is illustrated in 
Figure~\ref{fig:fields_pmarrow} as the green arrow. 
We find that these three stars have no obvious association with any 
of the known streams in this figure. However, a closer inspection 
reveals that there is a hint of a horizontal structure that roughly 
runs at DEC$\simeq 6$\degr. This may be a stellar stream 
undiscovered so far due to the Sgr Stream being too bright 
in this part of the sky. We note that this may also be an artifact 
created during the image processing stage. A deeper imaging 
survey in this region of the sky such as the Pan-STARRS 
or LSST may be able to reveal whether this structure is an actual stellar 
stream or not. For now, we tentatively call this the ``Parallel 
stream,'' since it runs parallel to the Celestial Equator.
If this is a stellar stream, its progenitor is likely a disrupted 
dwarf galaxy instead of a star cluster based on the fact that the 
1-d tangential velocity dispersion is $20.9 \kms$ (see 
Table~\ref{tab:pmavgs}, comparable to those of the Orphan and Sgr 
stream stars in this study.

\subsection{Field {\it ORPH-F4}: Probable Association with the Lethe Stream}

Three stars are found in the {\it ORPH-F4} field that likely belong 
to a group of stars moving in the same direction. The reflex solar 
motion at the distance of 12 kpc in this direction of the sky is found 
to be $(\mu_{\rm W}, \mu_{\rm N}) = (1.41, -3.50)~\masyr$, so the 
resulting net 2-d motion of the three stars in this field become $(\mu_{\rm W}, 
\mu_{\rm N}) = (-0.41, 1.91)~\masyr$. This vector of motion is illustrated in 
Figure~\ref{fig:fields_pmarrow} as the orange arrow.
Unlike the case of the three stars in the {\it ORPH-F3} field, it is 
immediately apparent that the three stars in this field likely belong 
to the Lethe stream, another stellar stream discovered by \cite{gri09}.
The Lethe stream is found to show a distance gradient decreasing from 
east to west: within the SDSS northern footprint used by \cite{gri09}, 
the distance at the eastern end is 13.4 kpc and at the western end near 
the Sgr stream, the distance decreases to 12.2 kpc. Our distance of 12 kpc 
is therefore consistent with the distance gradient found by 
\cite{gri09} for the Lethe stream. The three stars we found in this field 
also share the same stellar population properties (metallicity and age) 
as the Lethe stream as found by \cite{gri09}. Finally, \cite{gri09} 
concluded that Lethe is the debris stream of a globular cluster, 
and our measured 1-d tangential velocity dispersion of $5.6 \kms$ is 
consistent with this. We therefore conclude that the 3 stars belong to 
the Lethe stream.

\section{Summary and Conclusions}
\label{sec:conclusions}

We present high precision PMs of stars identified in four fields selected 
along the Orphan Stream using \hst. The combination of PM data and 
photometry has allowed for unique association of groups of stars with 
similar ages, spatial location, and kinematics. In the {\it ORPH-F1} field, 
we identified five stars that likely belong to the Orphan stream based on 
their stellar population, distance, and PM information. The average PM 
generally agrees with the model prediction by \citet{new10}. 
In the {\it ORPH-F2} field, we have a tentative detection of one star 
that belong to the Orphan stream. In the remaining two fields, we were not 
able to identify stars that belong to the Orphan stream. 

We also serendipitously identified and measured PMs of stars that belong 
to two known and one unknown stellar streams. In the {\it ORPH-F1} field, 
we detected three potential candidate stars that may belong to the Sgr 
faint arm based on their stellar population and 2-d motions on the sky.
However, our comparison to $N$-body models of \citet{law10} shows that 
only one of the three stars has the correct PM for being a member of the 
Sgr Stream. In the {\it ORPH-F2} field, we measured the PMs 
of seven stars that belong to the leading arm of the Sgr Stream. The average 
measured PM is consistent with the \citet{law10} model, just as we 
found in our Sgr Stream study \citep{soh15}. 
In the {\it ORPH-F3} field, we identified three stars that likely 
belong to a new stream at the distance of $\sim 32$ kpc, which 
we tentatively named the Parallel Stream.  
The stellar population and 1-d tangential velocity dispersion 
suggest that the progenitor is a dwarf galaxy.
Finally, in the {\it ORPH-F4} field, we measured for the first time 
the PMs of 3 stars that belong to the Lethe stream \citep{gri09}. 
The distance, stellar populations, net 2-d motion 
on the sky, and 1-d tangential velocity dispersion of the 
three stars are all consistent with the Lethe stream as their origin.

Finally, we found that fields {\it ORPH-F1} and {\it ORPH-F4}
show significant excess in their observed PM diagrams when compared 
to the Besan\c{c}on model. This indicates that there are likely 
additional substructures in these parts of the sky, but due to our 
limited field of view, we were not able to identify them in our study.
A deep and wide photometric or astrometric survey (e.g., Gaia) of 
these areas may reveal unknown substructures.

As part of an ongoing \hst\ archival legacy program (AR-13272, 
PI: R.~P.~van der Marel) to determine PMs of metal-poor halo stars 
in random pointings multiply imaged by \hst, we expect further PM 
measurements of stars that belong to known and unknown stellar streams. 

Once its final catalog is released, the {\it Gaia} mission will enable  
mapping of the entire extent of each stellar stream as well as the 
discovery of new stellar streams in the Galactic halo. However, for 
distances beyond $\sim 10$ kpc, only \hst\ is able to measure useful PMs 
for old stellar populations on a star-by-star basis. This study has 
illustrated the power of combining the high accuracy PMs and deep 
photometry allowed by \hst\ to uniquely identify dynamically associated 
groups of stars. As such, \hst\ will continue to be greatly advantageous 
and complementary to the {\it Gaia} mission in this research area.

\acknowledgments

We thank the anonymous referee for constructive feedback that helped
improve the presentation of our results.
Support for this work was provided by NASA through a grant for program
GO-13443 from the Space Telescope Science Institute (STScI), which is
operated by the Association of Universities for Research in Astronomy
(AURA), Inc., under NASA contract NAS5-26555. SRM acknowledges support 
from grant NSF-AST1312863.

\facility{Facilities: HST(ACS/WFC)}.

\end{document}